\documentclass[journal=nalefd,manuscript=letter]{achemso}
\setkeys{acs}{keywords = true}

\usepackage[T1]{fontenc} % Use type 1 (vector) fonts
\usepackage{epsfig}
\usepackage{amsmath}

\title{
Single-Layered Hittorf's Phosphorus: A Wide-Bandgap High Mobility 2D Material  \\
}
 \author{Georg Schusteritsch} \email{gs550@cam.ac.uk}
\affiliation{Department of Physics and Astronomy, University College London, Gower Street, London WC1E 6BT, United Kingdom
}
 \altaffiliation{
Current address: Department of Materials Science and Metallurgy, University of Cambridge, 27 Charles Babbage Road, Cambridge CB3 0FS, United Kingdom
}

 \author{Martin Uhrin }
 \altaffiliation{
Current address: Theory and Simulation of Materials (THEOS), and National Centre for Computational Design and Discovery of Novel Materials (MARVEL), \'Ecole Polytechnique F\'ed\'erale de Lausanne, CH-1015 Lausanne, Switzerland
}
\affiliation{Department of Physics and Astronomy, University College London, Gower Street, London WC1E 6BT, United Kingdom
}

 \author{Chris J. Pickard } 
\affiliation{Department of Physics and Astronomy, University College London, Gower Street, London WC1E 6BT, United Kingdom
}
 \altaffiliation{
Current address: Department of Materials Science and Metallurgy, University of Cambridge, 27 Charles Babbage Road, Cambridge CB3 0FS, United Kingdom and Advanced Institute for Materials Research, Tohoku University
2-1-1 Katahira, Aoba, Sendai, 980-8577, Japan
}

\date{\today}


\begin{document}

%\begin{tocentry}
%\includegraphics[height=3.5cm]{fig_abstract_graphic_a.png}
%\includegraphics[height=3.5cm]{fig_abstract_graphic_b.eps}
%\end{tocentry}

\begin{abstract}
We propose here a two-dimensional material based on a single layer of violet or Hittorf's phosphorus. Using first-principles density functional theory, we find it to be energetically very stable, comparable to other previously proposed single-layered phosphorus structures. It requires only a small energetic cost of approximately $0.04~\text{eV/atom}$ to be created from its bulk structure, Hittorf's phosphorus,  or a binding energy of $0.3-0.4~\text{J/m}^2$ per layer, suggesting the possibility of exfoliation in experiments. We find single-layered Hittorf's phosphorus to be a wide band gap semiconductor with a direct band gap of approximately $2.5$~eV and our calculations show it is expected to have a high and highly anisotropic hole mobility with an upper bound lying between $3000-7000$~cm$^2$V$^{-1}$s$^{-1}$. These combined properties make single-layered Hittorf's phosphorus a very good candidate for future applications in a wide variety of technologies, in particular for high frequency electronics, and optoelectronic devices operating in the low wavelength blue color range.
 \end{abstract}

The last decade, since the experimental realization of large-scale graphene\cite{Novoselov2004}, has seen an explosion of the field of two-dimensional materials. Although graphene offers very high mobilities, in excess of $10^5~\text{cm}^2\text{V}^{-1}\text{s}^{-1}$ if no charged impurities nor ripples are present~\cite{Morozov2008}, it is a semimetal with zero band gap in its ideal structure and it is not trivial to turn it into a semiconductor with a bandgap~\cite{Schwierz2010}. Graphene, although a candidate for many future technologies, is therefore still limited in terms of its possible electronics and optoelectronics applications. To address these shortcomings, much effort has since been focused on two-dimensional transition metal dichalcogenides, which inherently exhibit band gaps. They are however found to have relatively low mobilities: Even the highest mobility results reported for instance for MoS$_2$~\cite{Radisavljevic2011} or WSe$_2$~\cite{Chuang2014} have been found to be around only $200~\text{cm}^2\text{V}^{-1}\text{s}^{-1}$. Many electronics applications would highly benefit from a direct, wide-bandgap material with a higher mobility~\cite{Fiori2014, Xia2014a, Akinwande2014}, and the quest for a material fulfilling all these requirements is still ongoing.

Recently, single-layered black phosphorus, henceforth called black phosphorene, was studied theoretically, and found to have high and highly anisotropic hole mobility, of order of $10,000~\text{cm}^2\text{V}^{-1}\text{s}^{-1}$, and a direct band gap of approximately $1.5$~eV~\cite{Qiao2014}. Experiments on single-layered black phosphorene have shown mobilities of $286~\text{cm}^2\text{V}^{-1}\text{s}^{-1}$~\cite{Ye2014}, while using several monolayers have yielded hole mobilities of at least $1000~\text{cm}^2\text{V}^{-1}\text{s}^{-1}$~\cite{Li2014e}. Though very desirable for many applications and widely studied theoretically and experimentally~\cite{Fiori2014, Xia2014a, Akinwande2014}, it would be advantageous for many electronics and optoelectronics applications to find a two-dimensional material with a larger, direct, band gap. One may not have to look very far, since phosphorus exists in nature in many metastable phases~\cite{Bachhuber2014}. 

At higher pressures black phosphorus undergoes a transition to the A7 phase~\cite{Jamieson1963}, a single layer of which, henceforth called A7 phosphorene, was recently proposed as blue phosphorene~\cite{Zhu2014}. Using first-principles methods it was found to be a semiconductor with an indirect band gap of up to $2.9$~eV, however, it is not clear if A7 phosphorene can be easily formed experimentally.  In addition to black and white, a family of up to six phases of so called red phosphorus have been suggested~\cite{VanWazer1958,Stephenson1969}.  Form I is amorphous and can be formed from the controlled heating of white phosphorus.  Further heating in molten lead produces forms II and III which are not well characterized and van Wazer suggests~\cite{VanWazer1958} that these may be poorly crystallized versions of form IV, known as Hittorf's (or violet) phosphorus~\cite{Thurn1969}.  Form V, known as fibrous red phosphorus~\cite{Ruck2005}, is also crystalline and in DFT calculations close in energy to Hittorf's~\cite{Bachhuber2014} while form VI is only stable at liquid nitrogen temperatures~\cite{VanWazer1958}.  Both Hittorf's and fibrous red phosphorus are composed of layered connected chains of clusters and are known to be stable at room temperature.

In this Letter we focus on the single layer form of Hittorf's phosphorus, which, in allusion to the other \textit{X}enes, we suggest calling either monolayer Hittorf's phosphorus, Hittorf's phosphorene, or simply "hittorfene" in order to distinguish it from black phosphorene, which is often just called "phosphorene". We find that hittorfene can be expected to exfoliate based on energetic arguments: It has a lower binding energy per area per layer than black or A7 phosphorene. We calculate it to have a hole mobility with an upper bound in the range of $3000-7000~\text{cm}^2\text{V}^{-1}\text{s}^{-1}$, an order of magnitude higher than the highest mobility theoretically and experimentally known two-dimensional transition metal dichalcogenides~\cite{Radisavljevic2011, Chuang2014, Zhang2014c} and similar to the theoretical predictions made for black phosphorene~\cite{Qiao2014}. Hittorfene has, in comparison to black phosphorene, a significantly larger direct band gap of approximately $2.5$~eV, close in magnitude to that of the indirect band gap of A7 phosphorene. This combination of properties - a large direct band gap and high mobility - may make it an excellent candidate for technological applications: For instance electronics applications where any high frequency device requires high mobility materials and large on/off current ratios, and optoelectronics applications for which neither two-dimensional transition metal dichalcogenides nor black phosphorene can currently provide large enough direct band gaps to access the blue color range.

All our calculations are performed using density functional theory (DFT) as implemented in the CASTEP (version 16.1)~\cite{Clark2005} plane-wave code. Structural relaxations are performed using the limited-memory Broyden-Fletcher-Goldfarb-Shanno (L-BFGS) method.  The exact XC energy is approximated by the PBE generalized gradient approximation (GGA)~\cite{PBE_1996_PRL}. We perform calculations that account for dispersion corrections using the Tkatchenko and Scheffler (TS)~\cite{Tkatchenko2009} and Grimme (G06)~\cite{Grimme2006} approach. Higher level calculations to determine the band gap are performed using the hybrid Heyd-Scuseria-Ernzerhof functional (HSE06) ~\cite{Heyd2006,Heyd2003} and based on the relevant GGA structures.  We use on-the-fly-generated core-corrected ultra-soft pseudopotentials~\cite{Lejaeghere2016} that treat the valence electrons for the 3s$^2$, 3p$^3$ states for P for all calculations except our HSE calculations for which we employ norm-conserving pseudopotentials with the same valency. A plane wave cutoff energy of  $500~\text{eV}$ was chosen. We use a Monkhorst-Pack grid density with a maximum distance between two \textit{k}-points of better than $2\pi \times 0.03 \text{~\AA}^{-1}$, appropriately scaled for the calculations of the two-dimensional materials. All single-layered calculations are performed using vacuum surrounding the sheet of more than $13$~{\AA} but tested for convergence up to $20$~{\AA}. We assess the mobility by means of a phonon limited scattering approach.  This has been successfully applied to study  graphene~\cite{Xi2012, Fiori2013, Chen2013a, Bruzzone2011}, and various other two-dimensional materials, such as h-BN~\cite{Bruzzone2011}, two-dimensional dichalcogenides~\cite{Zhang2014c} and black phosphorene~\cite{MorganStewart2015a, Fei2014a, Qiao2014}. The approach is based on the deformation potential theory of Bardeen and Shockley~\cite{Bardeen1950} and approximates the electron-phonon coupling  by electron scattering of longitudinal acoustic phonons. The latter are modelled by uniform lattice dilations. We take the effective mass approximation to simplify the expression of the mobility, which then for a two-dimensional systems gives an expression for the mobility along direction $\alpha$ and for band $i$ of, ~\cite{Xi2012}

 \begin{equation}\label{Eq:2D:mobility}
\mu_{\alpha}^{\left(i\right)}=\frac{e \hbar^3 C_{\alpha}}{k_B T m_{\alpha}^* m_d \left(E_{1 \alpha}^{\left( i \right)}\right)^2}, 
\end{equation}

where $k_B$ is the Boltzmann constant, $T$ the temperature and $m_{\alpha}$, $C_{\alpha}$ and $E_{1 \alpha}^{\left( i \right)}$ are the effective mass, elastic modulus and deformation potential for $\alpha=x,y$. We take here the temperature to be $T=300~\text{K}$ as is common in similar studies~\cite{Xi2012, Chen2013a, Fei2014a, Qiao2014}. As in previous work the average effective mass was calculated as $m_d=\left( m_x^* m_y^*\right)^{1/2}$. The deformation potential, $E_{1 \alpha}^{\left( i \right)}$ is defined by $E_{1 \alpha}^{\left( i \right)}= \frac{\Delta E_i}{\Delta l_{\alpha}/l_{\alpha}^{\left(0\right)}}$, where $\Delta E_i$ is the change of the energy of the $i$'th band, $l_{\alpha}^{\left(0\right)}$ is the equilibrium lattice constant and $\Delta l_{\alpha}$ is the change of the lattice constant along $\alpha=x,y$ on compression or dilation. We consider here the conduction and valence band for the $i$'th band.  We compress and dilate the lattice in steps of $0.5\%$. The elastic moduli along $\alpha=x,y$ are determined by considering that $C_{\alpha} \frac{\Delta l_{\alpha}}{l_{\alpha}^{\left(0\right)}} = \frac{E-E_0}{S_0}$, where $E$ is the total energy for compressed/dilated structures, $E_0$ the total energy at equilibrium, and $S_0$ is the equilibrium lattice area in the {\it xy}-plane. The effective mass is found by fitting the band structure at the conduction band minimum (CBM) and valence band maximum (VBM)  with the nearly free electron model. This approach to assess the mobility only considers the electron - acoustic phonon limited coupling and hence should only be considered as an upper limit as other processes such as scattering due to optical phonons, structural defects and impurities may affect the mobility. However it has been shown to be a valuable approach to give indicative values of the order of magnitude and possible anisotropy in the mobilities.

\begin{figure}
\center{
\includegraphics[width= 16 cm]{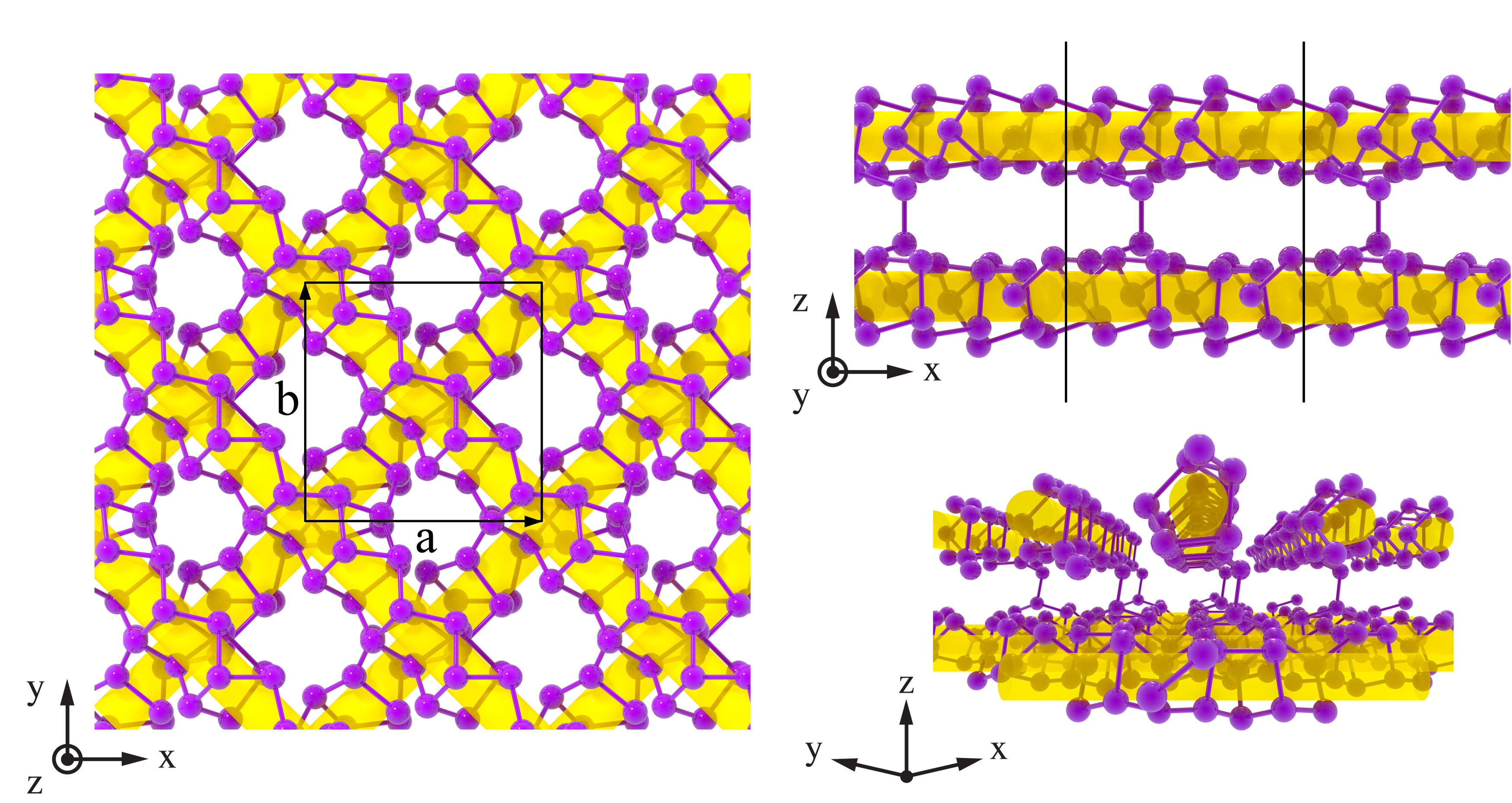}

\caption{
Crystal structure of hittorfene. Violet spheres represent phosphorus atoms, whilst the yellow tubes are added as guides to the eye only to emphasize the fibrous nature of hittorfene. The unit cell is shown as black solid lines for the orthographic top and side views.
}
\label{Fig:hittorfene:struct}
}
\end{figure}

Violet or Hittorf's phosphorus consists of fibers making up two sets of layers each consisting of two bonded levels of fibers (see Supporting Information)~\cite{Thurn1969}. To create hittorfene we cut the structure of Hittorf's phosphorus such that one covalently bonded layer remains. As such it is the thinnest possible layer that can be formed from Hittorf's phosphorus by only cutting van der Waals bonded layers and leaving all covalent bonds intact. The crystal structure of hittorfene is shown in Fig.~\ref{Fig:hittorfene:struct}. It consists of fibers cross-hatched in two levels with a total of 42 atoms in the unit cell. We have added yellow semi-transparent tubes into the structures shown in Fig.~\ref{Fig:hittorfene:struct} to emphasize the fibrous nature of hittorfene and Hittorf's phosphorus. The yellow tubes were chosen to fit with maximal diameter to not intersect the phosphorus structure. The fibers within each level are not connected, but instead the bonding occurs across them. The lattice constant for hittorfene is found to be $a =9.25~{\text\AA}$ and $b =9.26~{\text\AA}$ with $\gamma =90^{\circ}$ and an average distance between the two levels of fibers of $d =6.10~{\text\AA}$ using PBE without dispersion corrections. This changes only slightly when TS and G06 corrections are included, giving $a^{\text{TS}} =9.15~{\text\AA}$, $b^{\text{TS}} =9.22~{\text\AA}$, $d^{\text{TS}} =5.82~{\text\AA}$ and $a^{\text{G06}} =9.15~{\text\AA}$, $b^{\text{G06}} =9.21~{\text\AA}$, $d^{\text{G06}}  =5.78~{\text\AA}$, respectively. For completeness we perform our main results both with PBE and also PBE+TS to ensure the validity of our main conclusions.

\begin{figure}
\center{
\includegraphics[height=9.cm]{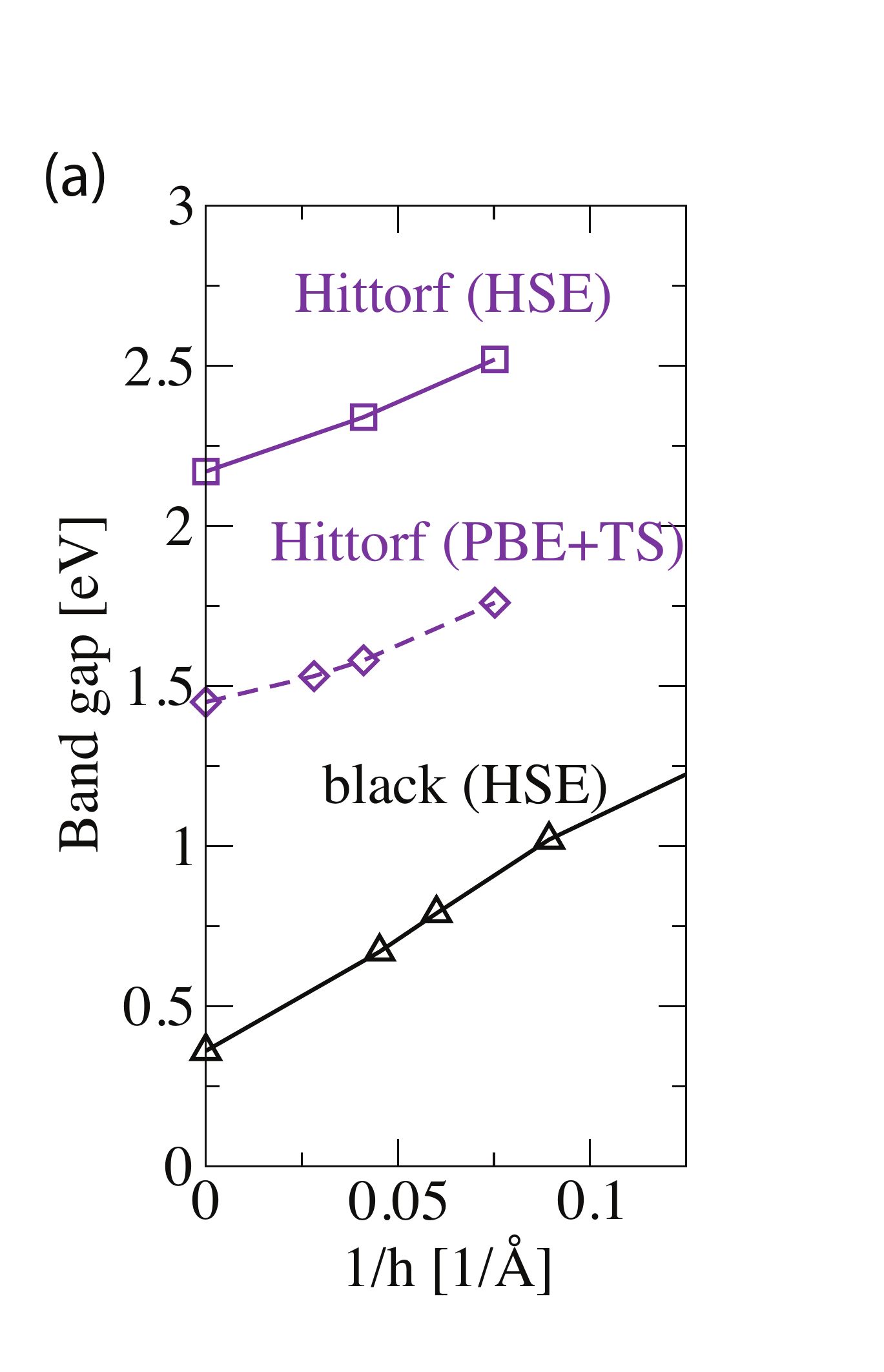}
\includegraphics[height=9.cm]{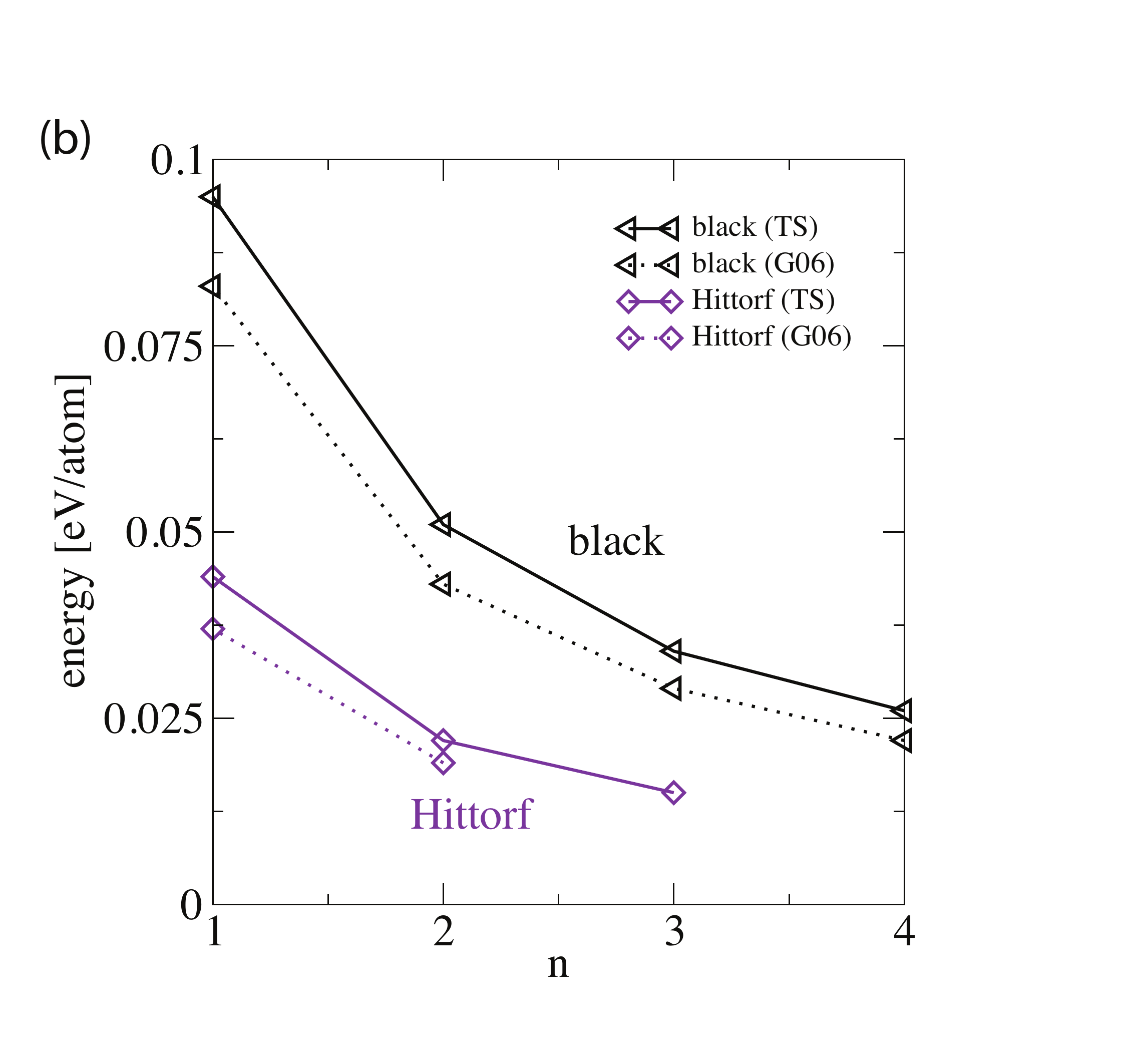}
\caption{
a) Band gap of Hittorf's and black phosphorus as a function of thickness. Both HSE and PBE+TS results are shown for Hittorf's phosphorus. The HSE data of black phosphorus shown are based on the band gap results of Qiao et al.~\cite{Qiao2014} in combination with their respective thickness taken from our own PBE+TS results. b) Relative energy of Hittorf's and black phosphorus with $n$ layers with respect to their bulk forms. For $n$=1 this becomes the binding energy of a single layer per atom. Solid and dotted lines mark results based on PBE+TS and PBE+G06, respectively. Results for $n$ layers of Hittorf's and black phosphorus are labelled as "Hittorf" and "black", respectively.
}
\label{Fig:bandgap:stability}
}
\end{figure}

We first consider the energetics of single and multi-layered Hittorf's phosphorus and compare it to those of black and A7 phosphorus. Fig.~\ref{Fig:bandgap:stability} (b) shows the relative energy of $n$-layers of hittorfene with respect to bulk Hittorf's phosphorus as violet diamond shapes for $n=1-3$. The binding energy increases monotonically with decreasing n, reaching $0.044~\text{eV/atom}$ ($0.037~\text{eV/atom}$) for PBE+TS (PBE+G06) calculations of $n=1$ (hittorfene). We show in the same figure the binding energy for $n=1-4$ layers of black phosphorus, which reaches a much larger value of $0.095~\text{eV/atom}$ ($0.083~\text{eV/atom}$) for PBE+TS (PBE+G06) calculations for black phosphorene (n=1). The latter compares well with the value found using the same van der Waals correction method by Sansone et al. ($0.084-0.089~\text{eV/atom}$)~\cite{Sansone2015}. Since we consider 2-D materials of different thicknesses it is helpful to calculate the binding energy per layer per area, that is the energy to completely separate or exfoliate the bulk into single layers: For hittorfene we find a value of $0.35~\text{J/m}^2$ ($0.29~\text{J/m}^2$) for PBE+TS (PBE+G06), smaller than the value for black phosphorene of $0.40~\text{J/m}^2$ ($0.35~\text{J/m}^2$) or A7 phosphorene of $0.42~\text{J/m}^2$ ($0.41~\text{J/m}^2$). In order to assess the propensity for exfoliation it is helpful to compare these values to the value of the binding energy of separating graphite into graphene, which was determined to be $0.56~\text{eV/atom}$ or approximately $0.34~\text{J/m}^2$ using Quantum Monte Carlo calculations~\cite{Spanu2009}. Other theoretical and experimental approaches give similar order of magnitude results~\cite{Wang2015}. This strongly suggests exfoliation of hittorfene should be possible.

\begin{figure}
\center{
\includegraphics[width=8.cm]{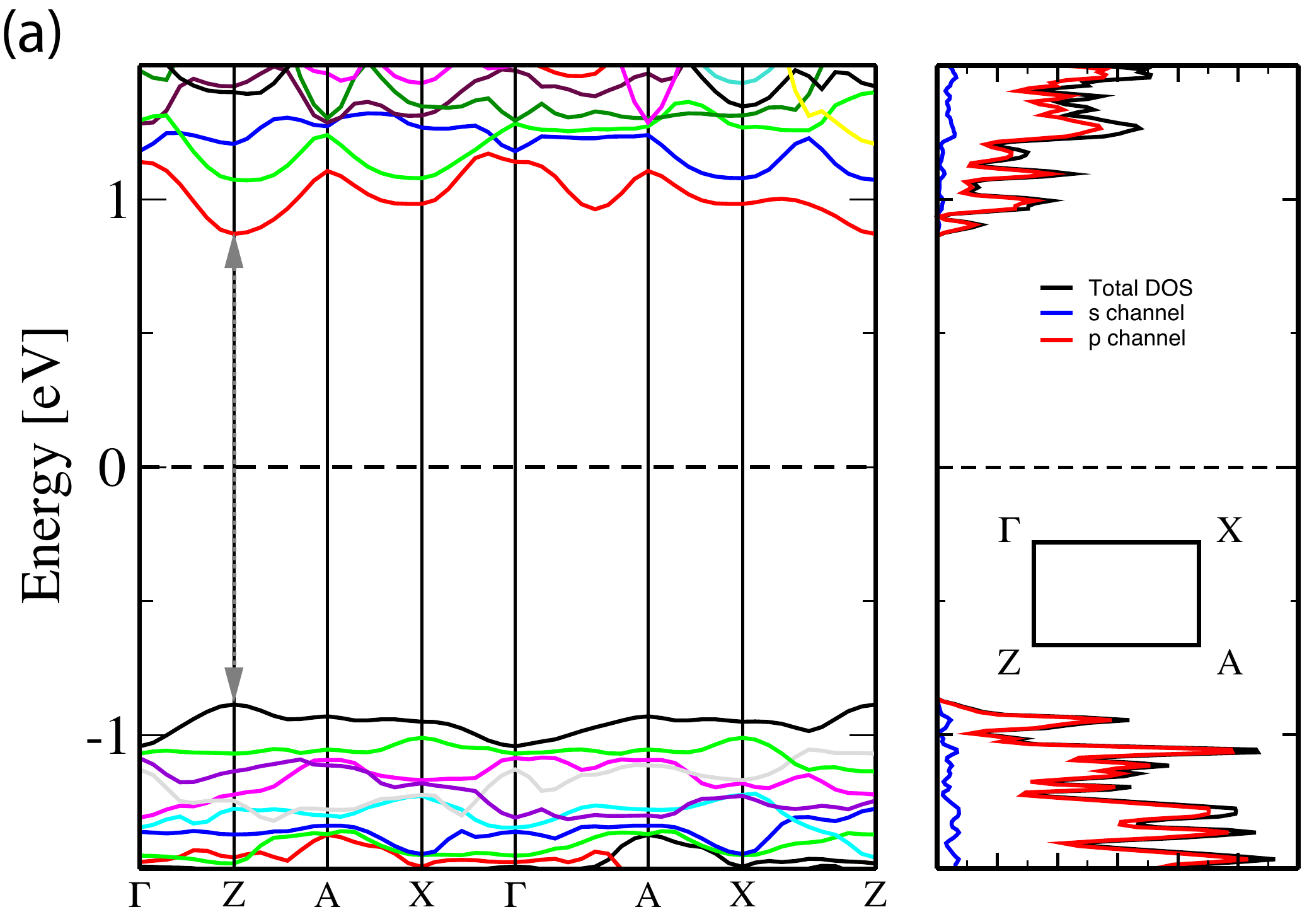}
\includegraphics[width=8.cm]{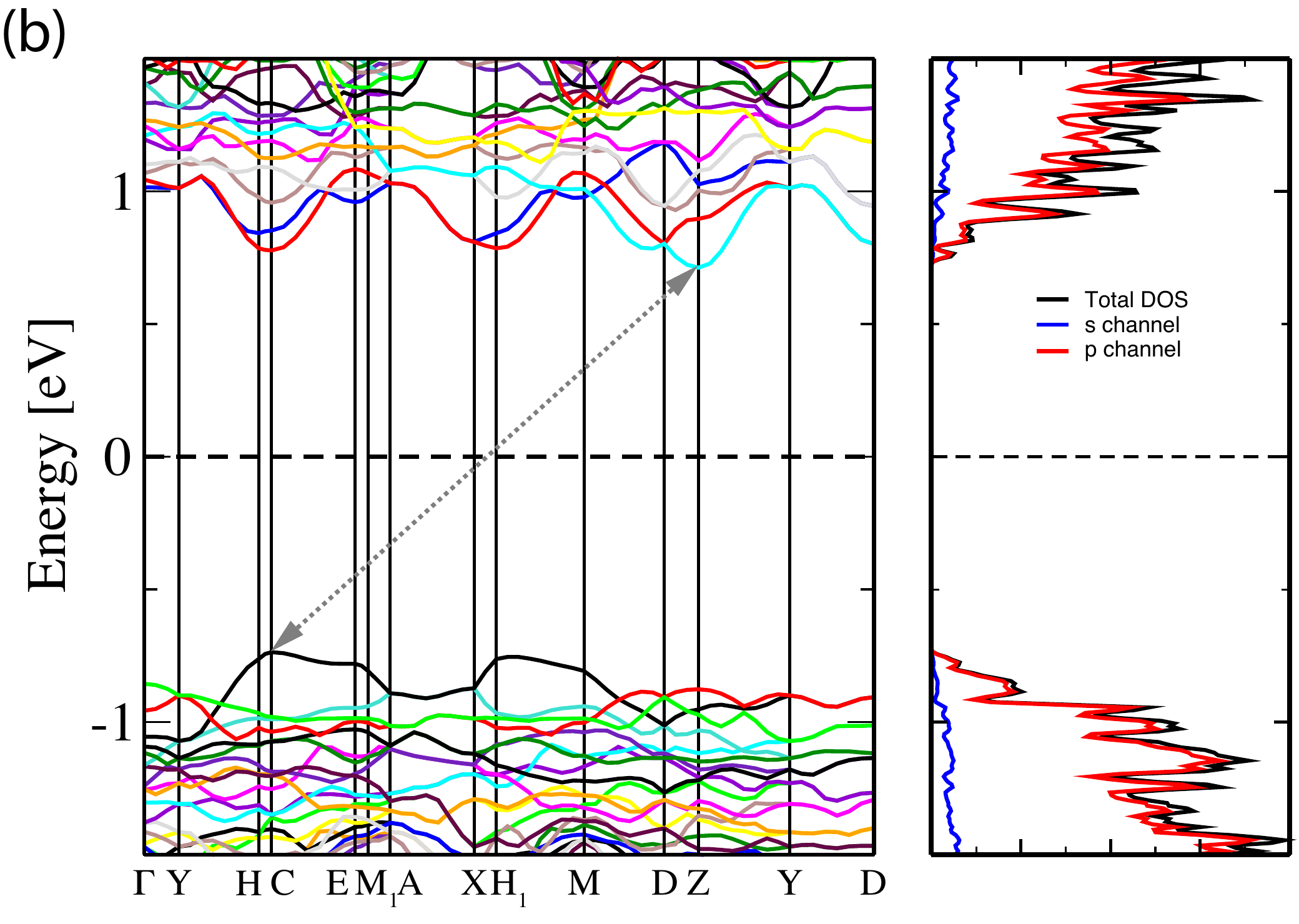}
\caption{
Band structure, total density of states and partial density of states of (a) hittorfene and (b) Hittorf's phosphorus based on PBE+TS calculations. A schematic of the reciprocal lattice with the high symmetry points indicated is shown for hittorfene as an inset in the plot of the density of states (see Supporting Information for further details on Hittorf's phosphorus) . Hittorfene exhibits a direct band gap at the Z point whilst Hittorf's phosphorus is found to have an indirect band gap in our calculations.
}
\label{Fig:hittorfene:bandstructure:DOS}
}
\end{figure}

%********************new table *******************

\begin{table}
 \begin{center}
  \begin{tabular*}{16.8cm}{@{\extracolsep{\fill}}l  c  c c c c }
  \hline \hline
  				&  \multicolumn{4}{c}{band gap [eV]}						&									\\
   				& PBE					& PBE+TS		&HSE 				&HSE (TS)			& type of band gap						\\	\hline
 Hittorf			& $1.82$					& $1.45$			& $2.62$ $(2.56)^{\left(c\right)}$					& $2.17$					& indirect $\text{C}\rightarrow\text{Z}$~$^{(*)}$									\\
 Hittorfene			& $1.93$					& $1.76$	[$1.72$]		& $2.71$					& $2.52$ [$2.47$]					& direct $\text{Z}\rightarrow \text{Z}$					\\
 Black			& $0.14$	 $(0.15)^{(a)}$		&  $0$~$^{(**)}$ 		& $0.68$ 					& $0.31$ $(0.36)^{(a)}$ 		& direct $\text{Y}\rightarrow \text{Y}$					\\
 Black phosphorene	& $0.90$ $(0.91)^{(a)}$		&  $0.84$			& $1.55$ $(1.53)^{(a)}$		& $1.48$						&direct $\Gamma\rightarrow\Gamma$		\\
 A7 phosphorene	& $1.93$ $(\sim 2)^{(b)}$		&  $1.93$			& $2.73$ $(\sim 2.9)^{(b)}$	& $2.74$						& indirect 								\\

  \hline \hline
  \end{tabular*}
  \caption{
Band gap and type of band gap of the different single-layered and bulk P allotropes using PBE, PBE+TS and HSE XC-functionals. Bulk A7 is found to be metallic hence not included here. Values in square brackets for hittorfene show the results based on G06 corrections. Round brackets are for comparison from various other theoretical work:$^{(a)}$: Ref.~\cite{Qiao2014}, $^{(b)}$: Ref.~\cite{Zhu2014}, $^{\left(c\right)}$: Ref.~\cite{Karttunen2007}. ~$^{(*)}$: We observe a complex behaviour of the band structure of bulk Hittorf's phosphorus on application of strain in the $z$-direction, for both the band gap and the location of the VBM and CBM. This results in an indirect bandgap for vdW corrected structures of bulk Hittorf's phosphorus, whilst PBE predicts a direct band gap (more details see Supporting Information). $^{(**)}$: We find that PBE+TS calculations show zero bandgap for bulk black phosphorus. 
   }
   \label{Table:band:gap}
 \end{center}
\end{table}

%********************new table end*******************

%************************new table*************************
\begin{table}
 \begin{center}
  \begin{tabular*}{16cm}{@{\extracolsep{\fill}}l  c  c  c c }
  \hline \hline
    				&  \multicolumn{2}{c}{PBE}						&		 \multicolumn{2}{c}{PBE+TS}								\\
 	 											&	electrons		& 	holes			&	electrons			& 	holes			\\	\hline
  $m_x^*/m_0$										&	$0.69\pm0.01$	&	$1.24\pm0.03$		&	$0.610\pm0.003$	&$1.41\pm0.01$		\\ 		
  $m_y^*/m_0$										&	$3.58\pm0.14$	&	$2.45\pm0.06$		&	$1.44\pm0.01$		&$1.65\pm0.03$		\\
  $E_{1x}$ [eV]										&	$1.40\pm0.03$	&	$1.26\pm0.06$		&	$1.50\pm0.02$		&$1.72\pm0.09$		\\
  $E_{1y}$ [eV]										& 	$0.66\pm0.01$	&	$0.18\pm0.05$		& 	$0.99\pm0.09$		&$0.37\pm0.05$		\\
  $C_x^{2{\text D}}$ [J/m$^2$]							& 	$49.70\pm0.2$	&	$49.70\pm0.2$		& 	$55.2\pm0.2$		&$55.2\pm0.2$			\\
  $C_y^{2{\text D}}$[J/m$^2$]							&	$49.92\pm0.4$	&	$49.92\pm0.4$		&	$54.8\pm0.28$		&$54.8\pm0.28$		\\
  $\mu_x^{2{\text D}}$ [$10^3$ cm$^2$V$^{-1}$s$^{-1}$]		&	$0.50\pm0.02$	&	$0.31\pm0.02$		&	$0.91\pm0.03$		&$0.18\pm0.02$		\\
  $\mu_y^{2{\text D}}$	[$10^3$ cm$^2$V$^{-1}$s$^{-1}$]	&	$0.43\pm0.03$	&	$7.68\pm4.3$		&	$0.88\pm0.16$		&$3.39\pm0.92$		\\			

  \hline \hline
  \end{tabular*}
  \caption{
Mobility $\mu_{\alpha}$, elastic modulus $C_{\alpha}$, deformation potential $E_{1\alpha}^{\left(i\right)}$ and effective mass $m_{\alpha} ^{*}$ for hittorfene employing PBE and PBE+TS.
   }
   \label{Table:mobility}
 \end{center}
\end{table}
%************************new table end*************************

To assess the applicability of hittorfene for future electronics applications, we next consider its band gap and compare it to that of the other known allotropes of phosphorus. We first calculate the band gap using the PBE XC-functional both with and without dispersion corrections, summarized in Table~\ref{Table:band:gap}. Bulk A7 is excluded from the table as it is found to be metallic. Comparing to known values of the band gap for the three different allotropes and their layered structures, shows consistent results with our calculations. We find hittorfene to have a larger band gap than bulk Hittorf's phosphorus, and a much larger band gap than either black phosphorus or black phosphorene. It is comparable to the band gap of A7 phosphorene, however it has a direct band gap, with transition at the Z point as shown in Fig.~\ref{Fig:hittorfene:bandstructure:DOS} (a). GGA functionals are known to underestimate the band gap; we therefore perform higher level theory hybrid exchange calculations using the HSE XC-functional to address this, summarized in Table~\ref{Table:band:gap} and Fig.~\ref{Fig:bandgap:stability} (a). The trends seen for PBE+TS persist once hybrid exchange is added to the calculations using the HSE functional to more accurately determine the band gap, and we find that hittorfene can be expected to have a wide band gap of approximately $2.5$~eV. To reproduce the experimental values for the band gap of bulk black phosphorus ($0.31-0.35~\text{eV}$~\cite{Warschauer1963, Maruyama1981, Akahama1983}) the structure used in the HSE calculations has to be dispersion corrected. Then a value of $0.31~\text{eV}$ is found in comparison to  $0.68~\text{eV}$ for a structure based on PBE, or a factor of more than two difference. The effect on hittorfene is found to be less significant: The band gap is different by less than 8\% if a TS corrected structure is used as the basis for the HSE calculations in comparison to PBE structures. The type of band gap, that is a direct band gap for hittorfene, is not affected by employing vdW corrections.

In contrast to black and A7 phosphorene the band gap of hittorfene is not as significantly larger than for its bulk structure. Although this behaviour may allow for tuning of the band gap over a larger spectrum in the case of black phosphorene, it at the same time also provides experiments with the challenge of having to very accurately select specific numbers of layers, as it is often difficult to have one layer only. This may not be as problematic for hittorfene as for black phosphorene, for which the band gap decreases rapidly with the number of layers~\cite{Qiao2014}, making experimental control more problematic. To investigate this further we consider the band gap of $n$ layers of hittorfene, plotted in Fig.~\ref{Fig:bandgap:stability} (a) as a function of the inverse of the width, $h$, of each sheet. We define $h$ here as the perpendicular distance between the outer-most atoms of each sheet and add to it a distance equal to twice the vdW radius~\cite{Bondi1964}. Within this more general definition of the width of a 2-D material, zero width sheets can be considered as well. This then shows a steady decrease of the band gap as the number of layers increases to bulk. This behaviour persists when high level theory HSE calculations are used to determine the value of the band gap. The overall behaviour is very similar when one considers the band gap of $n$-layers of black phosphorene (black curve in Fig.~\ref{Fig:bandgap:stability} (a)): black phosphorene has however a significantly smaller band gap for comparable sheet widths. 

To investigate the origin of the band gap increase from bulk Hittorf's phosphorus to hittorfene, we cleave ideal bulk Hittorf's phosphorus and calculate the band gap without any further ionic or volumetric relaxations. The value of $1.78~\text{eV}$ we find is nearly identical to that determined for a fully relaxed structure of hittorfene, thus showing that the band gap increase is a result of the lack of interlayer interactions. Furthermore by considering the PDOS it can be seen that the band extrema are dominated by p-type character.

We next consider the hole and electron mobilities that hittorfene exhibits. Our results are summarized in Table~\ref{Table:mobility} for both PBE and PBE+TS. We include the errors determined from fitting for the respective values; systematic errors may be larger. We first discuss our findings using PBE. The elastic moduli have a value of approximately $50~\text{J/m}^2$ in both the $x$ and $y$ directions. This lies approximately half way between the values for black phosphorene that has  highly anisotropic elastic moduli of $C_x=29~\text{J/m}^2$ and $C_y=102~\text{J/m}^2$~\cite{Qiao2014}. We find the effective mass of both the electrons and holes to be larger along the $y$ direction with the value of the electron effective mass along the $x$ direction being the smallest in magnitude. HSE calculations are found to be consistent with the values for the effective mass using PBE. Similarly to what was found for black phosphorene, there is a strong anisotropy of the deformation potential $E_{1\alpha}$, with that of $E_{1 y}^{\text {hole}}$ being the smallest in magnitude. This contributes strongly and results in a very large hole mobility along the $y$ direction of up to $7000~\text{cm}^2 \text{V}^{-1} \text{s}^{-1}$. The large error determined for this value is a direct result of the error propagation calculation involving the very small value of $E_{1y}$ and its associated error. It is also noteworthy that even the lower electron mobilities and hole mobility in the $x$-direction are similar order of magnitude but larger than the maximum values predicted for most two-dimensional dichalcogenides~\cite{Zhang2014c}. For comparison, these calculations are repeated using PBE+TS. The elastic constants remain nearly unchanged. The effective mass for both holes and electrons along the $x$-direction is also not strongly affected, whilst that along the $y$-direction does decrease more significantly. This has the effect of increasing the mobility for both holes and electrons in the $y$-direction. Similarly the deformation potential along the $x$-direction is not strongly affected, whilst that for the $y$-direction increases. This increase has the effect of decreasing the mobility. Due to the stronger (square) dependence of $E_{1\alpha}$, the combined effect results in a slight decrease of the final hole mobility along the $y$-direction, however lying within the error bounds of the value found for PBE without dispersion corrections. Given the general uncertainty of whether dispersion corrections are appropriate for studying hittorfene since no direct experimental comparison is possible yet, we cannot conclude which values should be considered more reliable. However both sets of calculations give results for the mobility that are of similar order of magnitude and the trends are the same. The differences in the values of the mobility between PBE and PBE+TS may hint at the possibility to control its value via strain engineering, which should be investigated in greater detail in future work.

In conclusion, we have studied a single layer of Hittorf's phosphorus. We find it to have a lower binding energy per layer per area in comparison to the two other known single-layered allotropes, black phosphorene, which has already been experimentally realized, and the theoretically predicted A7 phosphorene. It is only slightly higher in energy in comparison to bulk Hittorf's phosphorus, suggesting that it may be possible to exfoliate it experimentally. The mobility of hittorfene is found to be of similar order of magnitude as the predictions and measurements for black phosphorene: We find it to have a high and highly anisotropic hole mobility. The advantage over black phosphorene may however come in the form of a wide, direct band gap of up to $2.5~\text{eV}$. Hittorfene combines the advantages found for large bandgap materials, with the advantage of having a very high mobility as seen in black phosphorene, thereby making it an excellent candidate for next-generation device applications involving two-dimensional materials.

\begin{acknowledgement}
This work was supported in part by the EPSRC Grant EP/G007489/2 and EP/J010863/2. C. J. P. is supported by the Royal Society through a Royal Society Wolfson Research Merit award. All data supporting this study are provided as Supplemental Material accompanying this paper. Computational resources from the University College London and London Centre for Nanotechnology Computing Services as well as Archer as part of the UKCP consortium (EPSRC Grant EP/K013688/1) are gratefully acknowledged. This work was also partly performed using the Darwin Supercomputer of the University of Cambridge High Performance Computing Service (http://www.hpc.cam.ac.uk/), provided by Dell Inc. using Strategic Research Infrastructure Funding from the Higher Education Funding Council for England and funding from the Science and Technology Facilities Council.
\end{acknowledgement}

%\begin{suppinfo}
%The crystal structures of Hittorf's phosphorus and hittorfene are given in {\it cif} file format. Band structure, density of states and properties of bulk Hittorf's phosphorus as well the energetic stability of different phases of phosphorus is provided.
%\end{suppinfo}

\bibliography{hittorfene.bib}

\end{document}